\documentclass[a4paper,11pt]{article}
\pdfoutput=1 

\usepackage{jinstpub} 

\title{\boldmath On the Reflectivity of 128~nm Photons Off Stainless Steel}

\author[a]{S.L. Mufson\note{Corresponding author.}}

\affiliation[a]{Astronomy Department, Indiana University,USA}

\emailAdd{mufson@indiana.edu}

\abstract{
This investigation presents a new experimental determination of the reflectivity of 128~nm scintillation photons off stainless steel.  The experiment took place in the TallBo dewar facility at Fermilab. 
The data were obtained using a detector that is sensitive to photons falling on its front face but is insensitive to photons falling on its back face or its sides.  
By comparing the ratio of photons per track from tracks that illuminate the front face of the detector to those that cross the back side with a simulation of the experiment where the reflectivity can be varied, a value of the reflectivity was determined. 
The reflectivity for the SA-240-304L stainless steel alloy off which the scintillation photons reflect in the TallBo dewar as found by this experiment is R~$= 0.674 \pm 0.038$.  The experiment is not sensitive to the differences between specular and diffuse reflections.
}

\keywords{liquid argon scintillation, neutrino detectors, photon detection}

\arxivnumber{?} 

\begin{document}
\maketitle
\flushbottom

\section{Introduction}
\label{sec:intro}

Liquid argon (LAr) is currently being extensivly used as the active detector medium for many experiments that study weakly-intracting neutrinos and  search for massive weakly-intracting dark matter particles (WIMPs).  After interacting in the LAr, these weakly-interacting particles produce the charged and neutral daughters that provide crucial information central to the objectives of the experiments.  As the charged daughters traverse the LAr, they generate scintillation light that adds additional information to the  investigations.  Several innovative technologies are currently being developed to detect these scintillation photons~\cite{bib:howard,Abi:2020loh}

The assesment of the technologies for the detection of scintillation light typically requires an R\&D program that mounts the detectors in a test dewar.  Often the experiments themselves are staged in a moderately sized dewar.  The analysis of the data from these photon detectors, which frequently also uses a simulation of the experiment, usually requires the specification of the reflectivity of 128~nm scintillation photons off the stainless steel dewar walls.  The reflection of scintillation photons off stainless steel in LAr, however, is not well studied.  Currently, only one value of this parameter has been reported and it is found in an ICARUS internal report~\cite{bib:IcarusReflection} in a table without any supporting evidence.

This investigation presents a new experimental determination of the reflectivity of 128~nm scintillation photons off stainless steel.  The experiment took place in the TallBo dewar facility at Fermilab.  The run conditions and data collection for a parallel investigation that took place at the same time can be found elsewhere~\cite{bib:mufson-TallBo}.  

The experiment is straightforward in conception.  The data were obtained using an ArCLight detector~\cite{bib:ArCLight} that is sensitive to photons falling on its front face but is insensitive to photons falling on its back face or its sides.  Two sets of tracks were analyzed: (1) tracks that cross in front of and illuminate primarily the front face of the detector, so that most scintillation photons are viewed directly; and (2) tracks that pass on the back side of the detector, so that scintillation photons primarily reflected off the stainless steel walls of the dewar are detected.  By comparing the photons per track from track set  (1) to the photons per track from track set (2) with a simulation of the experiment where the reflectivity can be varied, a value of the reflectivity can be determined.  The simulation used for this determination has been previously described~\cite{bib:mufson-TallBo,bib:howard}.

\section{The TallBo Experiment}
\label{sec:TallBoExperiment}

The experiment took place in the liquid argon dewar facility TallBo, housed in the Proton Assembly Building (PAB) at Fermi National Accelerator Laboratory (FNAL).  The experiment ran from August 3, 2018 until October 17, 2018.  Track data from August 20, 2018 to September 9, 2018 were most appropriate for use in this analysis.   During these periods the ArCLight detector was exposed to both front side and back side tracks during uniform running conditions.  Details of the run conditions, including the levels of LAr purity that were present during these run dates, and the data quality can be found in~\cite{bib:mufson-TallBo}. 

\begin{figure}[h]
\centering
\includegraphics[width=4.5in,height=3.2in]{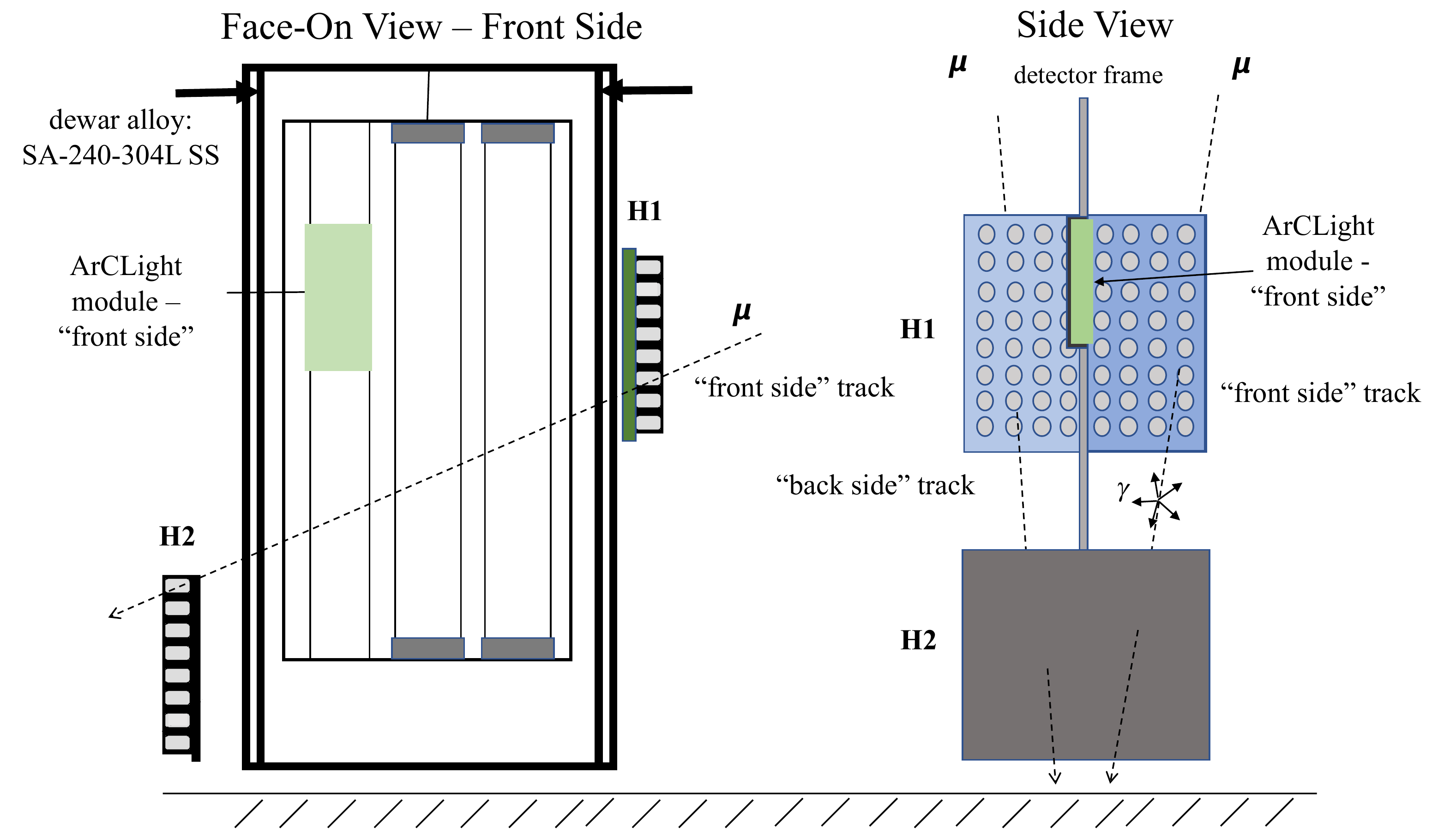}
\caption{Configuration of the experimental apparatus in the TallBo dewar.  As indicated, the dewar is construced of the stainless steel alloy SA-240-304L.  On the left is a face-on, front side view with a representative single cosmic muon track that triggers the readout.   On the right is a side view, rotated by 90$^\circ$ counterclockwise, with representative ``front side'' and ``back side'' muon tracks that trigger the readout.  Along the track scintillation photons are assumed to be emitted uniformly in solid angle.  Only photons striking the front side of the ArCLight are detected.   The back face and edges of the ArCLight are covered with a dielectric specular reflector foil (thickness exaggerated) that reflects away all scintillation photons generated in the LAr back into the LAr volume. }
\label{exptLayout}
\end{figure}

Fig.~\ref{exptLayout} shows two views of the experiment.
On the left is a face-on, front side view with a representative single cosmic muon track that triggers the readout.  As indicated, the stainless steel alloy out of which the dewar is made, and off which the scintillation photons are reflected, is  SA-240-304L.  The ArCLight photon detector~\cite{bib:ArCLight} shown in Fig.~\ref{exptLayout} collected the data for this experiment.  It was mounted in a custom frame that was suspended from the dewar lid and was read out by 4 Hamamatsu SiPM (MPPC) photodetectors addressed individually.  The ArCLight module was submerged uninterrupted in LAr during this phase of the experiment~\cite{bib:mufson-TallBo}.

On the right is a side view, rotated by 90$^\circ$ counterclockwise, with two representative muon tracks that  trigger the readout.  These tracks are wholly on one side of the detector frame or the other and are labelled ``front side'' if the tracks are viewed by the ArCLight directly or ``back side'' if they are not.  Tracks that cross from one side of the detector frame to the other were excluded from the analysis.  Along the tracks scintillation photons are assumed to be emitted uniformly in solid angle.  

For an event to trigger the DAQ, coincidence signals were required from the two hodoscope modules, H1 and H2, that flank the outside of the dewar, as pictured in Fig.~\ref{exptLayout}.  The DAQ trigger required a 4-fold coincidence: signals over threshold from both a PMT in H1 and in H2, and signals over threshold from both scintillator panels adjacent to H1 and H2, all signals were required to be detected within a 150~ns gate.  The thresholds and the 150~ns gate were set to reject large showers that would have overwhelmed the DAQ.  Background events during the hodoscope gate were typically of low energy.  A second cosmic ray event was unlikely during the gate.  Further details of the trigger can be found in~\cite{bib:mufson-TallBo}. 

\subsection{The ArCLight Detctor}
\label{ArCLight}

The ArCLight photon detector~\cite{bib:ArCLight} in Fig.~\ref{exptLayout} is designed to work by trapping and detecting 128~nm LAr scintillation photons that strike its front face.  The front face of the ArCLight is coated with a dichroic mirror film that is transparent in the blue and has high reflectance to other wavelengths striking it.   On the surface of this film is a layer of TPB that converts the VUV scintillation photons to blue photons, typically in the range 420 -- 450~nm.  These blue photons pass through the dichroic film and are efficiently absorbed by a 4 mm thick EJ280 WLS plate\footnote[1]{http://www.eljentechnology.com}. The back face and edges of the detector are covered with a dielectric specular reflector foil that reflects away all photons back into the LAr volume.  The waveshifted photons that pass into the ArCLight but are unabsorbed by the WLS plate strike this reflective foil and get directed back into the plate where they are then subsequently absorbed, thus increasing the efficiency of the ArCLight detector.  The photons trapped by the WLS plate are channeled by total internal reflection to its ends.  On one end, they are detected by 4 SiPMs, read out individually.  Since the opposite end of the plate is mirrored, the photons striking this end are reflected back toward the SiPMs.  Further details of the ArCLight detector can be found in~\cite{bib:ArCLight}.   

The experiment described here is feasible because only photons striking the front face of the ArCLight are detected while photons striking the back or sides are reflected away.  In particular, the signals from front side tracks are dominated by scintillation photons that directly illuminate the front side of the ArCLight.  
Conversely, the signals from back side tracks, detected on the front side, are dominated by scintillation photons that have been reflected off the the stainless stell dewar walls.  
Low energy background events detected during the 150~ns hodoscope gate also contribute to both front side and back side signals.   

The 4 individually addressed SiPMs that provide the ArCLight readout -- denoted: MMPC~0, MPPC~1, MMPC~2, and MPPC~3 -- were Hamamatsu 3x3mm MPPCs (S13360-3050VE\footnote[2]{https://www.hamamatsu.com/resources/pdf/ssd//s13360-2050ve$\_$etc$\_$kapd1053e.pdf}).  These MPPCs have 50$\mu$m pixels in a TSV package and are coated with epoxy resin.  Details of the laboratory measurements of the the performance characteristics of the similar 6x6mm MPPCs can be found in~\cite{bib:mufson-TallBo}.  It is expected that the performance characteristics of the 3x3mm devices is comparable to the 6x6mm devices.

\subsection{Operations}
\label{sec:operations}

The procedure used to fill the TallBo dewar with LAr is described in~\cite{bib:mufson-TallBo}.  When filled, the volume of LAr in TallBo was approximately 460~liters.  Once filled, the TallBo dewar was sealed and subsequently maintained at a positive internal pressure of 8 psig to prevent O$_2$, N$_2$, and H$_2$O contamination from the outside.  Gaseous argon from the ullage was recondensed to liquid argon and returned to the dewar after passing through a filter. 

The experiment consisted of recording the scintillation light from single cosmic muons traversing the LAr in the TallBo dewar defined by a four-fold coincidence trigger.  Two run periods are analyzed in this investigation.  One run included only front side tracks; the second included only back side tracks.  The dates of these runs are given in Table~\ref{tab:experimentalRuns}.  To avoid confusion, the nomenclature used to label the runs is the same as found in~\cite{bib:mufson-TallBo}. 
\begin{table}[h]
  \begin{center}
    \caption{Runs Analyzed}
    \vspace{0.2em}
    \label{tab:experimentalRuns}
    \begin{tabular}{| c |  c c  c  |}
      \hline
      \hline
     & Track &  Run & Run  \\
      Run        &  Geometry & Start  &  End  \\  
      \hline
       2 & front-side &  Aug 20, 2018 &  Aug 27, 2018  \\
      3& back-side &  Aug 28, 2018 & Sept 9, 2018   \\
         \hline
      \hline
   \end{tabular}
  \end{center}
\end{table} 

The event statistics are given in Table~\ref{tab:runStats}.  Column [2] gives the total number of hours of run time for the runs.  The number of four-fold coincidence triggers per second for each run are given in column [3].  Column [4] gives the number of four-fold coincidence triggers per second in which there was one and only one hit on a PMT in each hodoscope module (``Single Track Rate'').  These single track triggers were the ones analyzed in this investigation.  
\begin{table}[h]
  \begin{center}
    \caption{Run Statistics}
    \vspace{0.2em}
    \label{tab:runStats}
    \begin{tabular}{| c |  c c c |}
      \hline
      \hline
        & Run & 4-fold  & Single Track \\
     Run         & Time  &  Coinc. Rate  &  Rate \\  
             &  [hr] &  [Hz]  &  [Hz]\\  
      \hline
       2 & 333 & 0.062 & 0.015  \\
       3 &  203 & 0.056 &0.013\\
        \hline
      \hline
   \end{tabular}
  \end{center}
\end{table}
Table~\ref{tab:runStats} shows clearly that run rates for both four-fold coincidence triggers and single tracks were stable for both runs.  

\section{TallBo Simulation}
\label{sect:simulation}

The track simulation is described in~\cite{bib:howard,bib:mufson-TallBo}.  The simulation for single tracks assumes that cosmic muons in TallBo travel along straight paths with end points fixed at the centers of the two triggered PMTs in the hodoscopes on either side of the dewar.  In this simulation, the cosmic muons are all assumed to be minimum ionizing that create 40,000 scintillation photons/MeV~\cite{bib:howard,bib:scintYield2} = $8.42 \times 10^4$ photons/cm along their tracks.  Scintillation photons were generated uniformly along the track segment that passes through the LAr volume, with  the photon momentum vectors distributed uniformly in solid angle.   Once generated, photons were tracked along straight line paths until they intersected with the ArCLight or were lost.  Only photons that intersected with the front side of the ArCLight, which were tallied separately, were used in the analysis.  Along their tracks, photons could undergo a Rayleigh scattering, be absorbed by a contaminant (mostly N$_2$) along its path, reflect off the back side of the ArCLight, or reflect off the walls of the dewar~\cite{bib:howard}.  The reflectivity of the dewar walls and the fraction of light reflected, whether specular or diffuse, were input parameters to the simulation.  For specular reflections, the outgoing photon's direction was computed with respect to the normal to the dewar wall and the rotation of the polarization vector in the perpendicular plane was unchanged.  For diffuse reflections, the outgoing photon's direction was chosen to be uniform in solid angle and its polarization vector was uniform in the perpendicular plane.

Every track in the front side and back side TallBo data sets that struck two single hodoscope PMTs and passed the analysis cuts was simulated 10 times for each choice of reflectivity and specular/diffuse fraction.  
After all tracks in a data set were simulated, the resultant histograms were averaged to give the total mean number of photons striking the ArCLight for that choice of reflectivity and specular/diffuse fraction.  

\section{Data Analysis}
\label{dataAnalysis}

The front side (run~2) and back side (run~3) track samples selected for analysis were the "Single Tsrack" triggers in column [4] of Table~\ref{tab:runStats}.  Single tracks were excluded from the samples if a straight line between the PMT centers on H1 and H2 crossed the detector plane.  The tracks selected for analysis were assumed to be dominated by single minimum ionizing muons in liquid argon.  

The analysis began by filtering each waveform in the data samples with an 15-point running mean that was meant to smooth out fluctuations along the waveform.  For each smoothed waveform, the mean of the $\sim$2 $\mu$s of data before 
\begin{figure}[h]
\centering
\includegraphics[width=2.4in,height = 2.in]{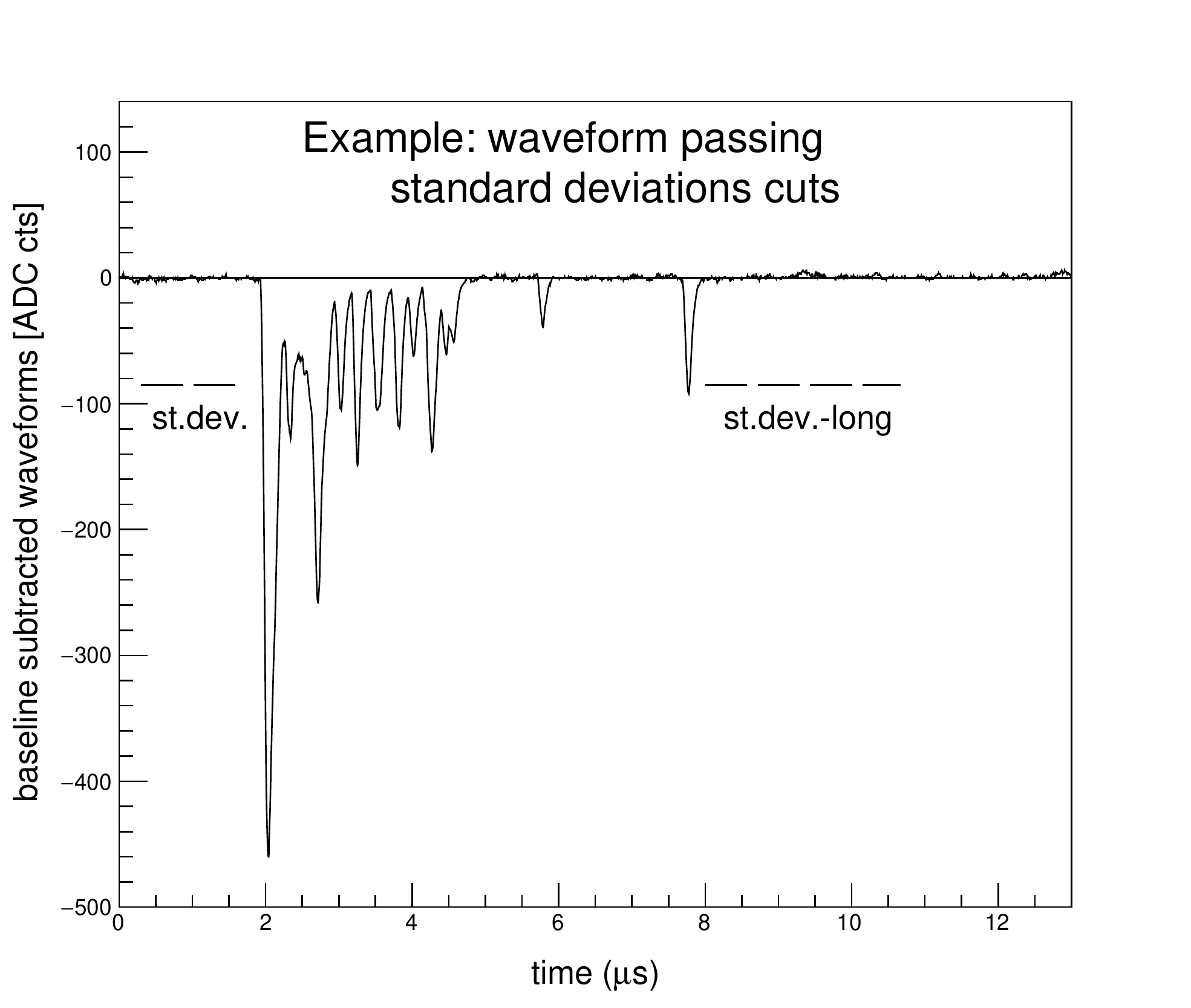}
\includegraphics[width=2.4in,height = 2.in]{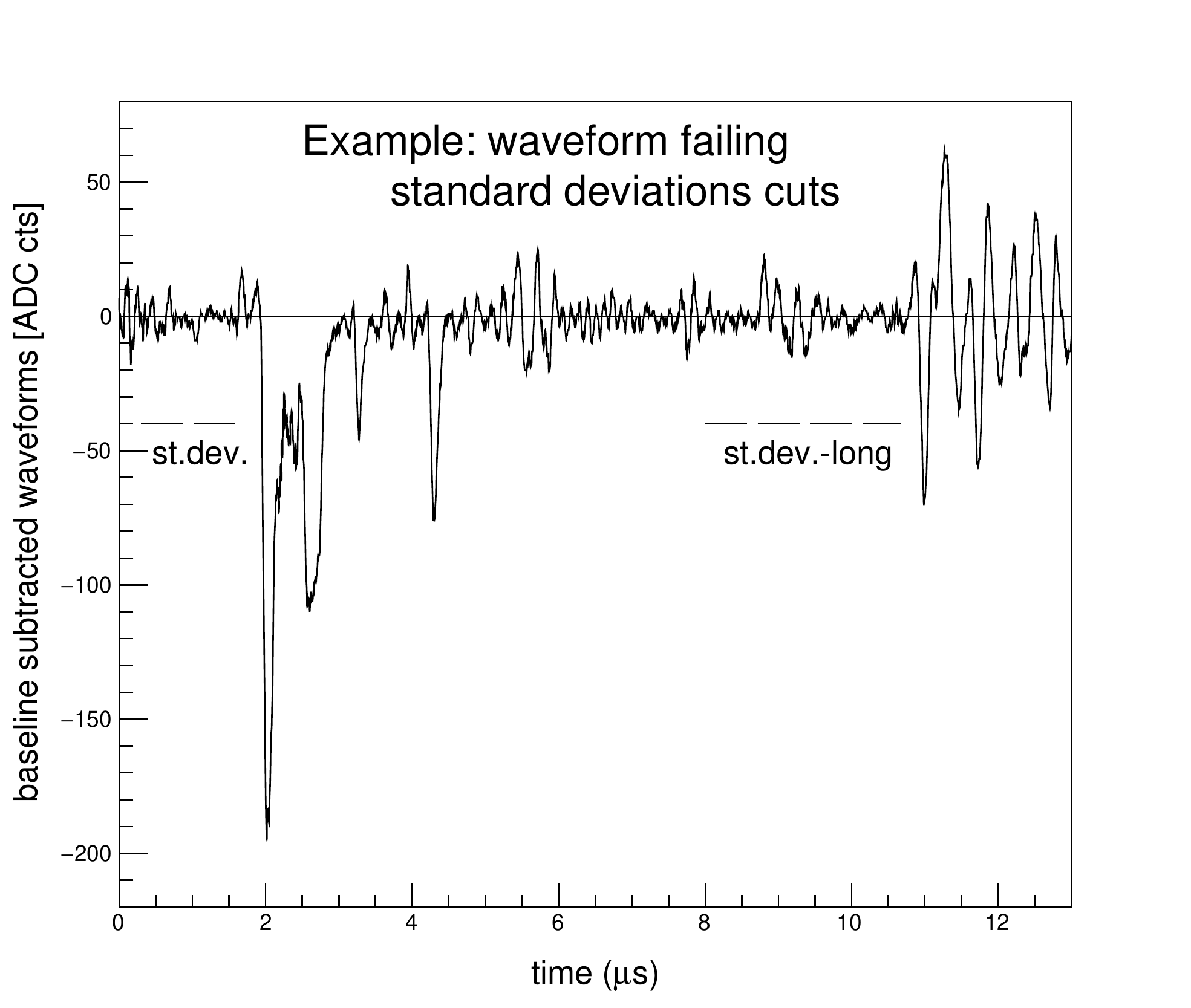}
\caption{Two representative waveforms from run~2.  The waveform segments over which the ``st.dev'' and ``st.dev-long'' cuts were implemented are indicated.  As seen in both waveforms, there are many individual photon hits after the  trigger.  On the {\it left} is a waveform that passes the cuts set at 3.0 standard deviations.  On the {\it right} is a waveform that fails these cuts.  The waveform on the {\it right} shows an example of the anomalous waveform behavior that Cut~(1), Cut~(2), and Cut~(3) were designed to remove.}
\label{waveforms}
\end{figure}
the trigger was defined as the baseline for that waveform.  Three cuts were applied to the  waveforms based on their baselines.  Cut (1) removed waveforms with large standard deviations in the $\sim$2 $\mu$s pretrigger data about the baseline (``st.dev'').  Cut (2) removed waveforms with large standard deviations about the baseline in the waveform from 8-10~$\mu$s  (``st.dev-long'').  Fig.~\ref{waveforms} shows where these cuts fall for two representative waveforms.  Cut (3) removed waveforms that had signifiant positive fluctuations ($>$ 15 ADC counts) between the trigger and 8~$\mu$s after the trigger.  These three cuts were implemented to remove anomalous waveforms with spurious shapes and/or unusually large fluctuations.  The right panel in Fig~.\ref{waveforms} shows an example of one such anomalous waveform.  These anomalous waveforms were found to correlate strongly with the 3 cuts that were found in a significant number of the run~2 and run~3 waveforms.  The origin of the anomalous behavior is likely to be electrical in nature but its source could not be determined and repaired during the experiment.  The time available between the arrival of the ArCLights in Bloomington and the scheduled start of the experiment at Fermilab was too short to adequately diagnose and repair the device before mounting it in the detector frame and shipping it.  

The values for Cut (1) and Cut (2) investigated were 2.5, 3.0, 3.5, 4.0, and 4.5 standard deviations.  The values for the st.dev and st.dev-long cuts in the analysis were set equal to the same number of standard deviations for each cut.  A cut more restrictive than 2.5 standard deviations was found to exclude too many waveforms for adequate statistics.  A cut less restrictive than 4.5 standard deviations was found to include too many anomalous waveforms in the analysis.  Although the track samples for the different cut values are clearly correlated, systematic trends in the results as more and more tracks with less restrictive cuts were added to the analysis did not prove to be significant.  

The waveforms for the selected tracks in the sample were integrated from the trigger out to 4$\times \tau_{\rm T} = 4.8 \mu$s, where $\tau_{\rm T} \approx 1.6\mu$s is the time constant for the de-excitation of the excited triplet state of $ \text{Ar}_2^* $ \cite{bib:TallBo,bib:N2Contamination} resulting in the ``late light'' in the LAr scintillation signal.  This integration limit includes $>$98$\%$ of the late light as the muon traverses the LAr.

\begin{figure}[h]
\centering
\includegraphics[width=2.4in,height = 2.in]{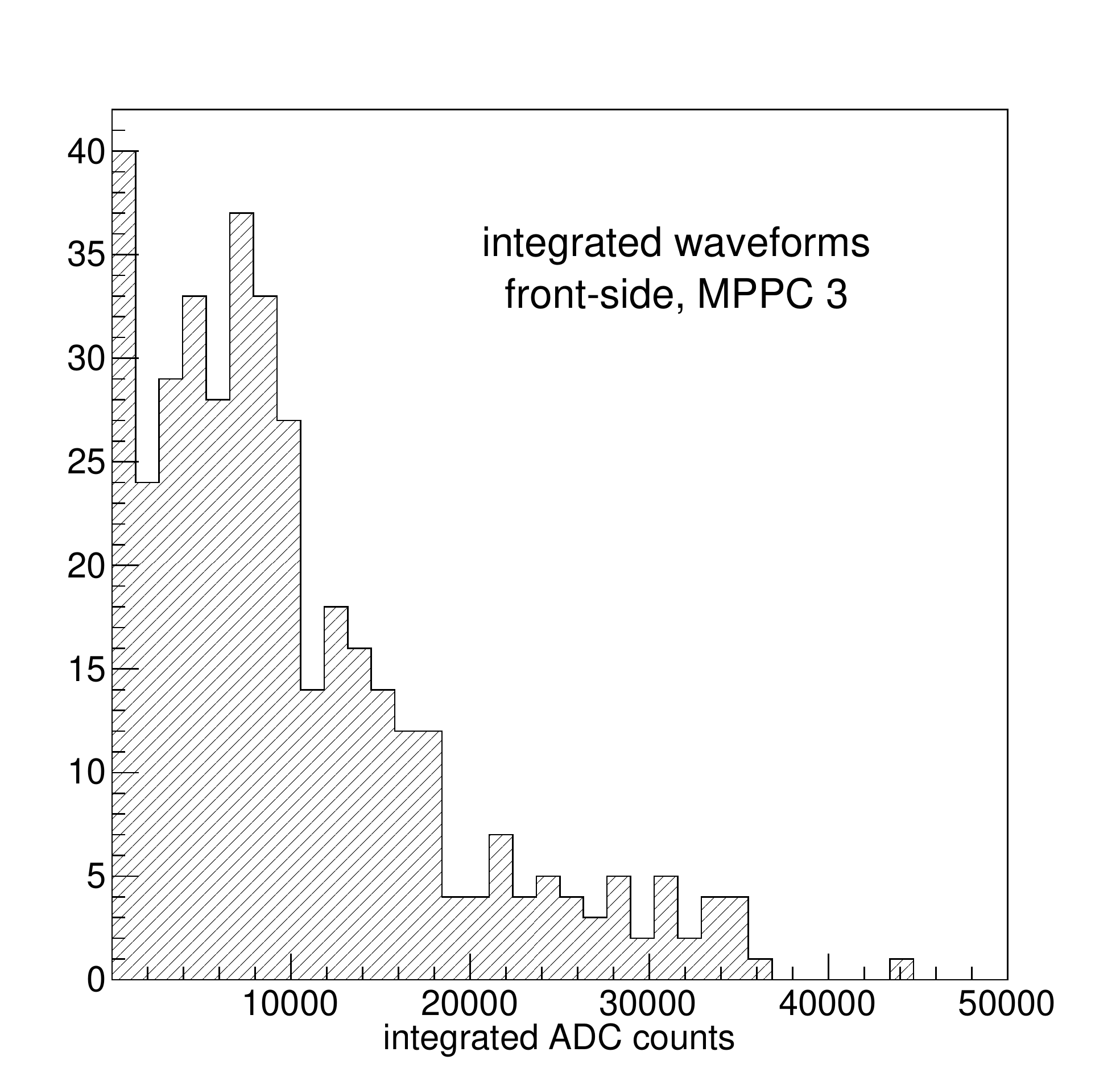}
\includegraphics[width=2.4in,height = 2.in]{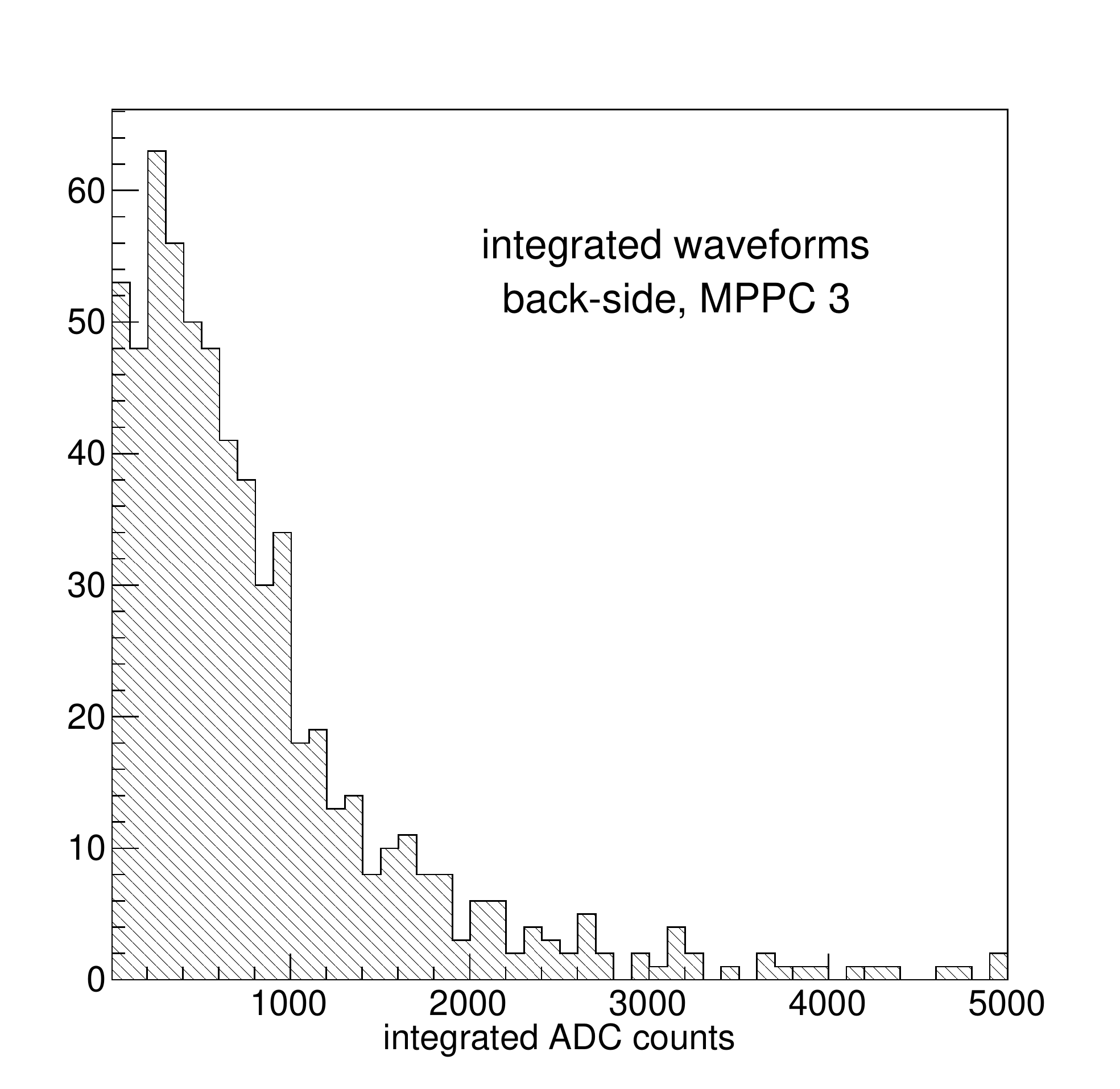}
\caption{The integrated waveforms from run~2 and run~3 for MPPC~2 for the st.dev and st.dev-long cuts set at 3 standard deviations.  Note that the horizontal scales of the histograms are 10x larger for the front side because the front side signals are significantly stronger.  The integrated waveforms in the main peak are associated with single, minimum ionizing cosmic muons.  The long tails are likely a mix of high energy muons and noise events.  The integrated waveforms in the lowest bin are low energy background events and were excluded from the analysis.}
\label{distributions}
\end{figure}

For the 5 choices of Cut~(1) and Cut~(2), the integrated waveforms for each of the 4 MPPCs were put into separate histograms, 5 for run~2 and 5 for run~3.  Examples of these histograms for MPPC~3 are shown in Fig.~\ref{distributions} for the st.dev and st.dev-long cuts set at 3 standard deviations.  Note that the horizontal scales of the histograms are 10x larger for the front side because the front side signals are significantly stronger. 
The main peak in these distributions of the integrated waveforms is associated with single, minimum ionizing cosmic muon tracks~\cite{bib:howard}.  The long tails are likely a mix of high energy muons and some noise events.  The integrated waveforms in the lowest bin are low energy background events~\cite{bib:howard}, not from single muon tracks, and were excluded from the analysis.  

\begin{figure}[h]
\centering
\includegraphics[width=2.4in,height = 2.in]{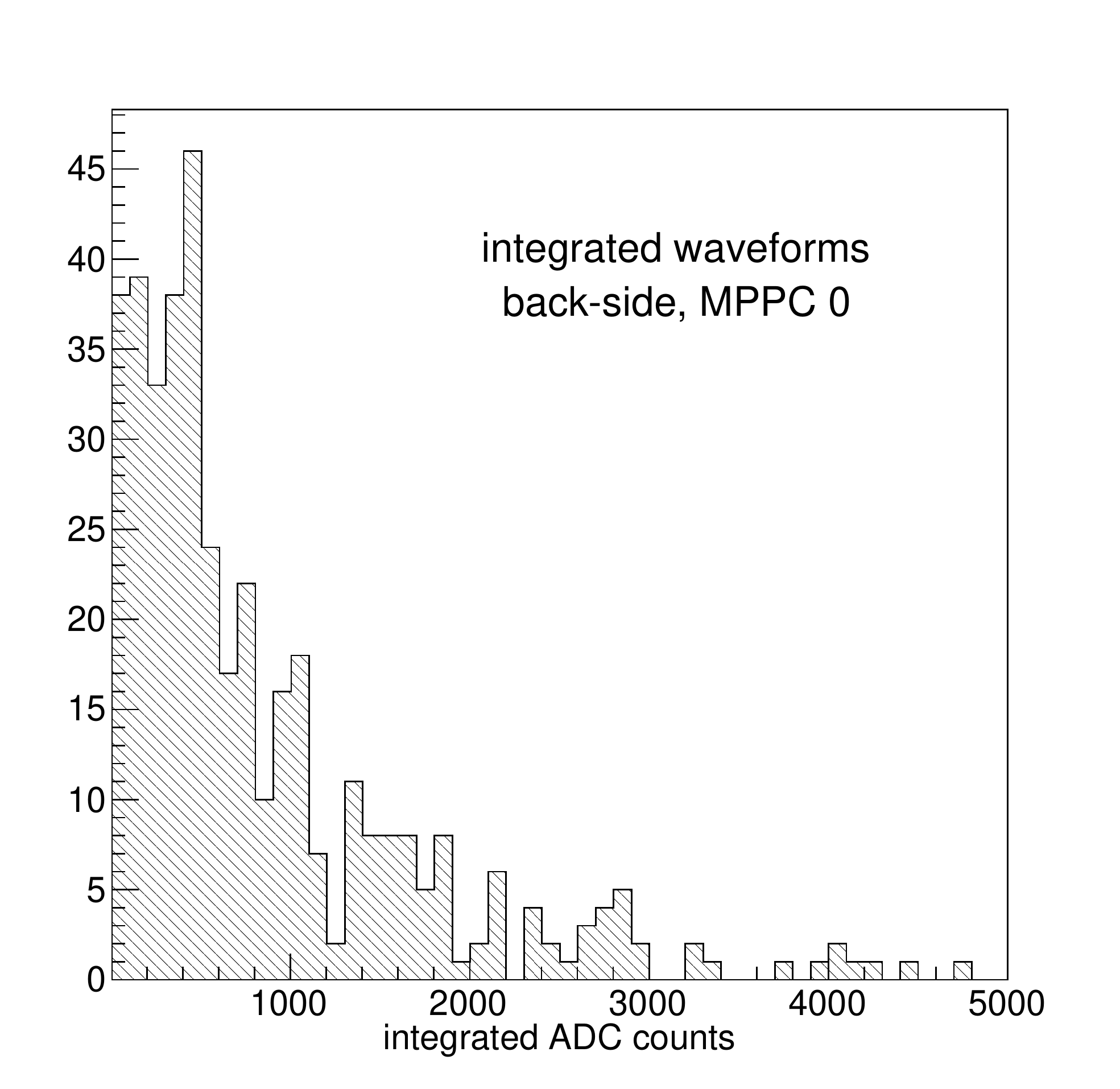}
\caption{The integrated waveforms from run~3 for MPPC~0 for the st.dev and st.dev-long cuts set at 3 standard deviations.  The integrated waveforms in the lowest two bins are low energy background events and were excluded from the analysis.}
\label{backSideMPPC0distribution}
\end{figure}

For the run~3 waveforms recorded by MPPC~0, the low energy background events are found in the lowest {\it two} bins, as can be seen in Fig.~\ref{backSideMPPC0distribution}.
This behavior is seen for all choices of Cut~(1) and Cut~(2).  The other MPPCs do not show this behavior in the lowest bins.  Fig.~\ref{backSideMPPC0distribution} suggests that MPPC~0 is more sensitive to low energy noise events and consequently records significantly more of them.  This behavior wuold not be seen in the run~2 histogram for MPPC~0 since the wider bin width includes these lowest energy bins in its first bin.  

The summed signal in the distributions for the single muons was computed by integrating the histogramss, excluding the lowest bin(s).  Since the lowest bins add very little signal, the results of the integrations are not sensitive to the precise lower limit of the integration.  However, the waveforms in these lowest bin(s) do  significantly affect the mean $\#$[ADC]/muon track, a key atatistic in this analysis, by adding waveforms with little signal.  

The results of summing the signals for the 4 MPPCs in run~2 and run~3 for the st.dev and st.dev-long cuts set at 3 standard deviations are given in Table~\ref{tab:ResultsStDev3}.
\begin{table}[ht]
\begin{center}
	\caption{Signal in Distributions of Integrated Waveforms for st.dev = st.dev-long = 3.0.}
	\vspace{0.2em}
	\label{tab:ResultsStDev3}
	\begin{tabular}{| c | c c c |}
		\hline
		\hline
 ArCLight &$\#$tracks &	 integrated signal & $\#$[ADC]/track\\
 MPPC & & [ADC] & \\  
	\hline
	\hline
\underline{(run 2)} & &  & \\  
MPPC 0  & 631 & $8.12\times10^6$ & $1.29\times10^4$\\  
MPPC 1  & 247 & $2.96\times10^6$ & $1.20\times10^4$\\
MPPC 2 & 410 & $4.61\times10^6$ & $1.12\times10^4$\\
MPPC 3 & 334 & $4.01\times10^6$ & $1.20\times10^4$\\  
\hline
\hline
\underline{(run 3)} & &  & \\  
MPPC 0 & 329 & $6.62\times10^5$ & $2.01\times10^3$\\  
MPPC 1 & 389 & $8.66\times10^5$ & $2.23\times10^3$\\
MPPC 2 & 604 & $1.25\times10^6$ & $2.07\times10^2$\\
MPPC 3 & 597 & $1.35\times10^5$ & $2.26\times10^3$\\  
\hline
\hline
	\end{tabular}
	\end{center}
\end{table}
In this table, column [1] lists the MPPC and run, column [2] enumerates the number of tracks from single muons in each distribution, column [3] gives the integrated signal in the distribution in [ADC] counts, and column [4] gives the $\#$[ADC]/muon track for that histogram.  Since each MPPC for the two runs contributes a different number of tracks to the integrated signal, the $\#$[ADC]/muon track is the  statistic most useful for comparing the signals in the different distributions.

In Table~\ref{tab:ResultsStDev3} the $\#$[ADC]/track for MPPC~2 for both run~2 and run~3 are low, suggesting that MPPC~2 is less sensitive than the other MPPCs.  The $\#$[ADC]/track for MPPC~0 is highest for run~2 front side tracks, while the $\#$[ADC]/track for the run~3 back side tracks is lowest.  One plausible explanation for this MPPC~0 anomaly is that it is as sensitive as MPPC~1 and MPPC~3 but, as discussed for Fig.~\ref{backSideMPPC0distribution}, there are more noise events in the MPPC~0 run~3 tracks.  This would increase the number of tracks without significantly increasing the signal, thereby lowering its  $\#$[ADC]/track for run~3.

Central to this investigation is the comparison of the signal from tracks directly illuminating the ArCLight in run~2 and tracks that are mostly seen by light reflected off the dewar walls in run~3.  This comparison can be effetively characterized by the ratio of the $\#$[ADC]/track in run~3 to the $\#$[ADC]/track in run~2.    
\begin{table}[ht]
\begin{center}
	\caption{Ratio of the $\#$ [ADC] counts/track in run~3 to the $\#$ [ADC] counts/track in run~2.}
	\vspace{0.2em}
	\label{tab:Results1}
	\begin{tabular}{| c ||  c | c | c | c | c | }
		\hline
		\hline
\multicolumn{1}{|c}{st.dev.\,cuts: } & \multicolumn{1}{c}{2.5} & \multicolumn{1}{c}{3.0} & \multicolumn{1}{c}{3.5} & \multicolumn{1}{c}{4.0} & \multicolumn{1}{c|}{4.5} \\
	\hline
         \hline
 \multicolumn{1}{|c||}{}& \multicolumn{5}{c|}{(run 3)/(run 2) = ([ADC]/track)$_{\rm run~3}$ $\div$ ( [ADC]/track)$_{\rm run~2}$} \\
        \hline
       \hline
MPPC 0 & 0.149 & 0.156 & 0.156 & 0.154 & 0.156  \\  
MPPC 1  & 0.216 & 0.186 &  0.182 & 0.191 & 0.188 \\
MPPC 2 & 0.156 & 0.184 & 0.203 & 0.199 & 0.201 \\
MPPC 3 & 0.146 & 0.188 &  0.192 &  0.202 & 0.192 \\  
\hline
\hline
<run3/run2> &0.167$\pm$0.033 &  0.179$\pm$0.015 & 0.183$\pm$0.020 & 0.186$\pm$0.022 & 0.184$\pm$0.020 \\  
\hline
\hline
	\end{tabular}
	\end{center}
\end{table}
Table~\ref{tab:Results1} gives these (run~3/run~2) ratios for the 4 MPPCs and for the 5 sets of analysis cuts.  The mean ratio for each analysis cut is given in the lowest row.  The error on the mean ratios is the standard deviation of the individual (run~3/run~2) ratios.  Table~\ref{tab:Results1} shows that the (run~3/run~2) ratios for Cut (1) = st.dev = Cut (2) = st.dev-long = 2.5 for MPPC~2 and MPPC~3 are not consistent with their (run~3/run~2) ratios for the remaining analysis cuts.  Further, the (run~3/run~2) ratio for MPPC~1 is inconsistent with the ratios for the other 3 MPPCs.  Finally, the mean <run3/run2> ratio differs for this cut and its standard deviation is significantly larger.  Since it is required that the analysis cuts be uniform for all the MPPCs in the analysis, the ratios for st.dev. cut = 2.5 were excluded from further analysis.  

\begin{figure}[h]
\centering
\includegraphics[width=3.5in,height=3.0in]{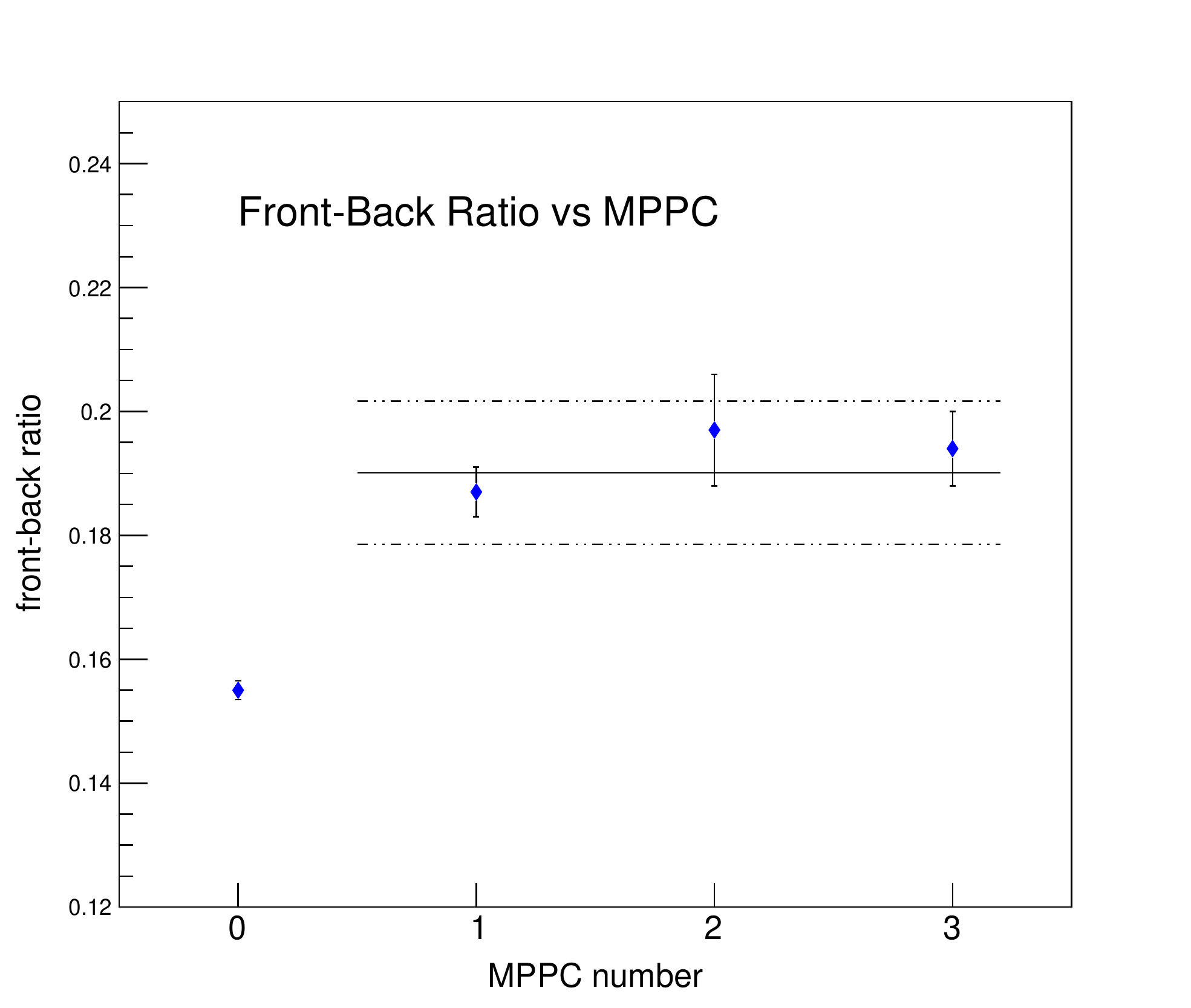}
\caption{Plotted are the mean ratios, <MPPC>, for each MPPC, as given in Table~\ref{tab:Results}.  The weighted mean of the <MPPC> are shown as a solid line.  The dashed line is offset from the weighted mean by the quadratic sum of the errors on the individual mean ratios.  The MPPC~1, MPPC~2, and MPPC~3 ratios are clearly consistent with one another, while the MPPC~0 ratio is an outlier.}
\label{front-backRatioMPPC}
\end{figure}

Table~\ref{tab:Results} gives the (run~3/run~2) ratios for the 4 MPPCs  for the remaining 4 sets of analysis cuts.
\begin{table}[ht]
\begin{center}
	\caption{Ratio of the $\#$ [ADC] counts/track in run~3 to the $\#$ [ADC] counts/track in run~2.}
	\vspace{0.2em}
	\label{tab:Results}
	\begin{tabular}{|  c || c | c | c | c ||c| }
		\hline
		\hline
\multicolumn{1}{|c}{st.dev.\,cuts: } & \multicolumn{1}{c}{3.0} & \multicolumn{1}{c}{3.5} & \multicolumn{1}{c}{4.0} & \multicolumn{1}{c}{4.5} & \multicolumn{1}{c|}{} \\ 
	\hline  
         \hline
\multicolumn{1}{|c||}{} &\multicolumn{4}{c||}{(run 3)/(run 2) = ([ADC]/track)$_{\rm run~3}$ $\div$ ( [ADC]/track)$_{\rm run~2}$}& <MPPC>\\
       \hline
\hline
MPPC 0 &0.156 & 0.156 & 0.154 & 0.156 & 0.154$\pm$0.003\\
\hline
MPPC 1  & ~~~~~0.186 ~~~~~&  ~~~~~0.182 ~~~~~& ~~~~~~0.191 ~~~~~& ~~~~~0.188 ~~~~~& 0.187$\pm$0.004\\
MPPC 2 & 0.184 & 0.203 & 0.199 & 0.201 & 0.197$\pm$0.009\\
MPPC 3  & 0.188 &  0.192 &  0.202 & 0.192 & 0.194$\pm$0.006\\  
\hline
\hline
	\end{tabular}
	\end{center}
\end{table}
In the last column, the mean ratio <MPPC> is given for each MPPC.  The error is the standard deviation of the individual ratios.  These ratios and their associated errors are plotted in Fig.~\ref{front-backRatioMPPC}.  
The weighted mean of the <MPPC> are shown as a solid line.  The dashed line is offset from the weighted mean by the quadratic sum of the errors on the individual mean ratios.  This figure shows clearly that MPPC~1, MPPC~2, and MPPC~3 ratios are consistent with one another, while the MPPC~0 ratio is an outlier.  The low (run~3/run~2) ratio for MPPC~0 is found for all analysis cuts.  The low value for the MPPC~0 ratio is likely related to a significant number of low energy noise events in run~3, as discussed previously.  Since its (run~3/run~2) ratios are inconsistent with the others, MPPC~0 was excluded from further analysis.

Table~\ref{tab:Results} shows that only $\sim$19$\%$ of the ADC counts in front side events are found in back side events.  
Since the tracks analyzed are all assumed to be minimium ionizing muons that generate a constant number of scintillation photons per cm as they traverse the dewar, the implication is that fewer scintillation photons from these tracks were detected from back side tracks. 
Fig.~\ref{waveforms} shows two illustrative waveforms chosen from the mid range of the distributions of integrated waveforms in Fig.~\ref{distributions} on the same vertical scale that indicate this hypothesis is plausible.  
\begin{figure}[h]
\centering
\includegraphics[width=2.4in,height = 2.in]{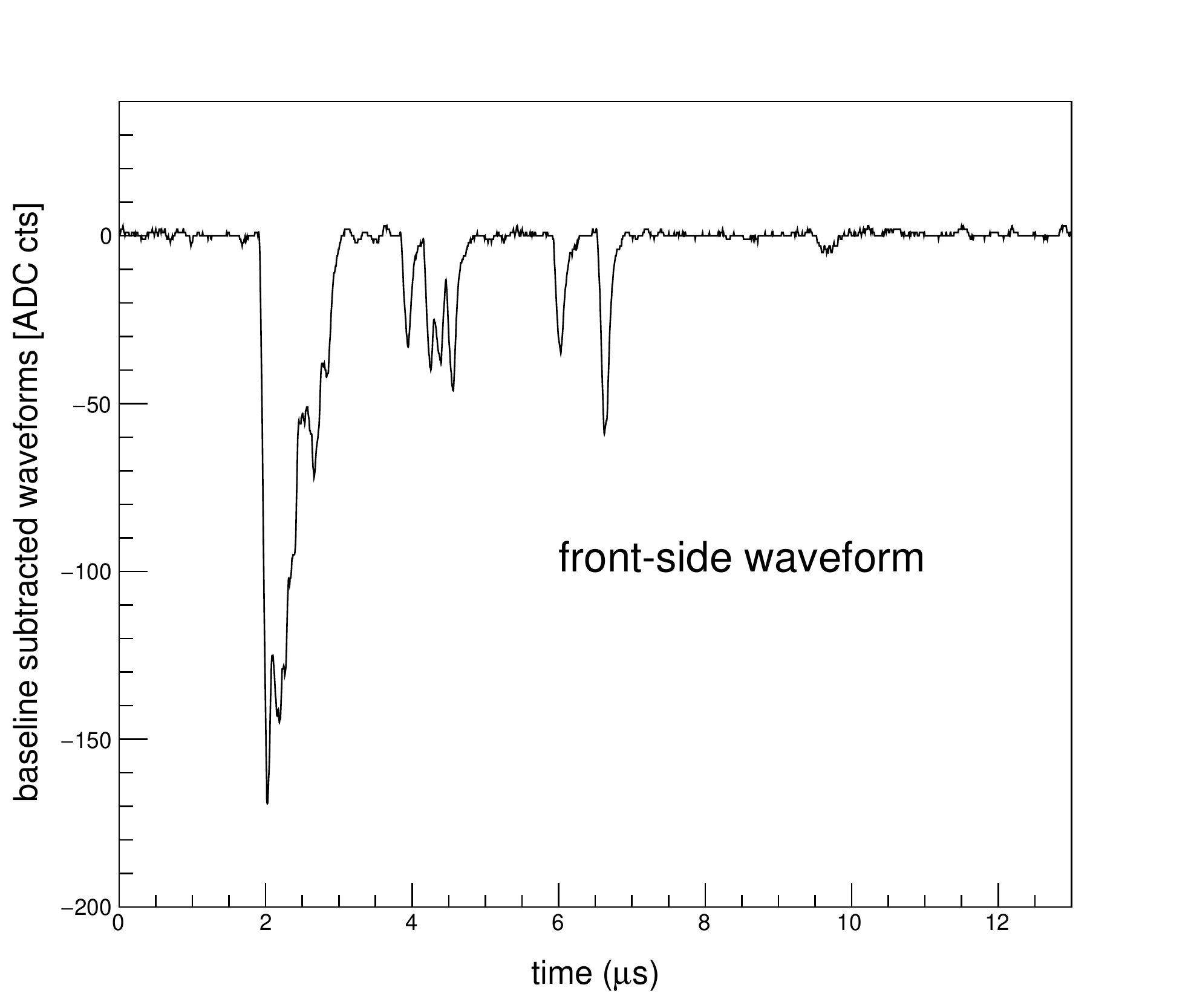}
\includegraphics[width=2.4in,height = 2.in]{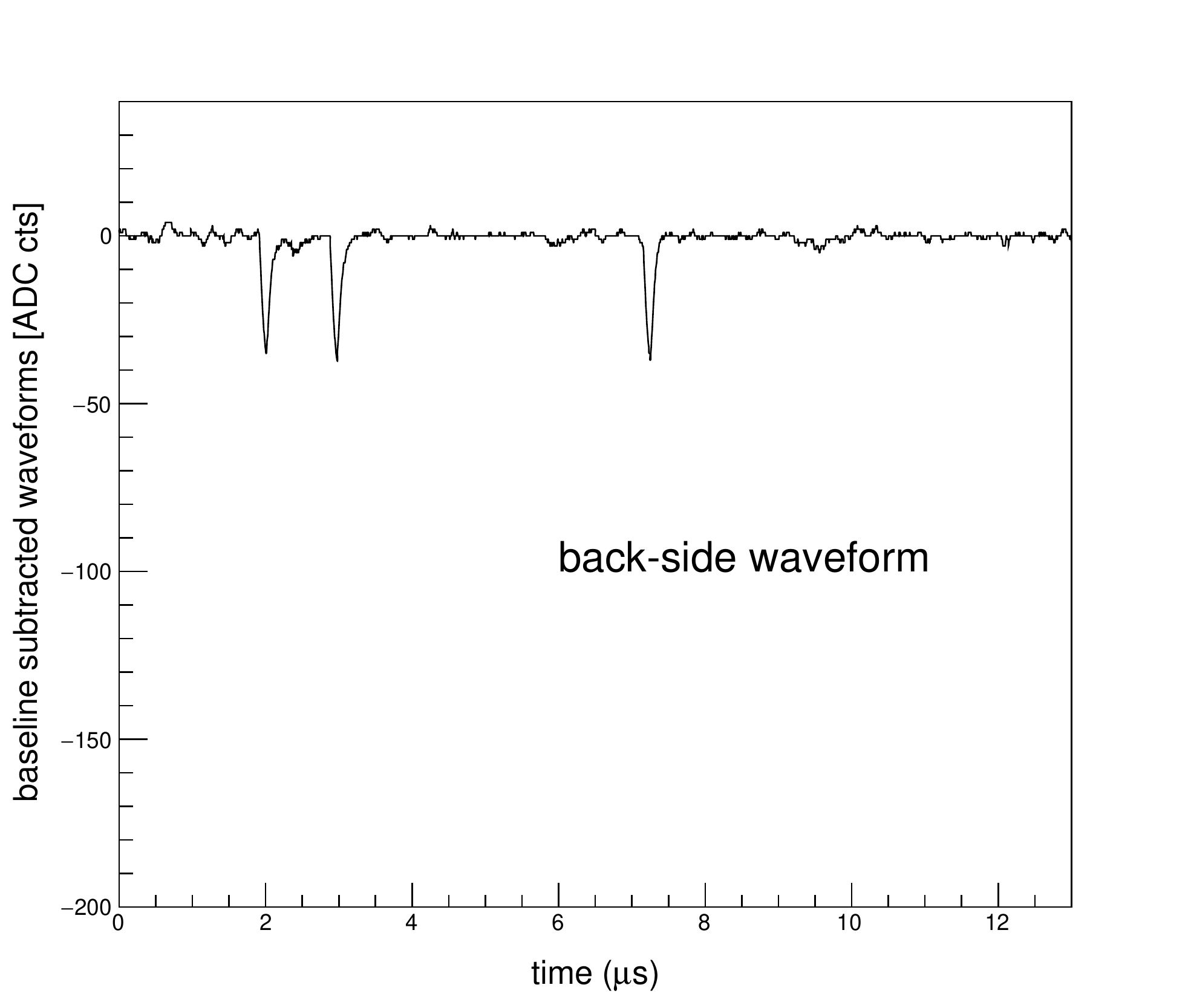}
\caption{Two illustrative waveforms chosen from the mid range of the distributions of integrated waveforms in Fig.~\ref{distributions} on the same vertical scale. These waveforms suggest that a significant fraction of the reflected photons are lost at the reflecting surface and accounts for the loss of scintillation photons in the back side tracks.}
\label{fig:waveforms}
\end{figure}
(The integral of the front side integrated waveform =  $15.5 \times 10^3$ ADC; the integral of the back side waveform = $1.8 \times 10^3$ ADC.)  The signal strength in the initial burst of early photons is plainly stronger in the front side waveform than in the back side waveform.
Also the late light signal in the back side waveform is also clearly suppressed.  These waveforms suggest that a significant fraction of the reflected photons are lost at the reflecting surface.

\section{Results}
\label{sec:results}

For each value of the (run3/run2) ratio for MPPC~1, MPPC~2, and MPPC~3 given in Table~\ref{tab:Results}, simulations were run with values of the dewar wall reflectivity varied between 0.0 and 1.0, and the ratio of specular to diffuse reflection set at 0.50.  For several values of the reflectivity, two addional simulations were run with the ratio of specular to diffuse reflection set at 0.25 or 0.75.  Each simulation for a particular MPPC and (run3/run2) ratio was computed for the specific set of tracks selected by Cut (1), Cut (2), and Cut (3).  Tracks in the lowest bin of the histograms of the integrated waveforms (e.g., Fig.~\ref{distributions}) were excluded.   

\begin{figure}[h]
\centering
\includegraphics[width=3.8in,height=3.2in]{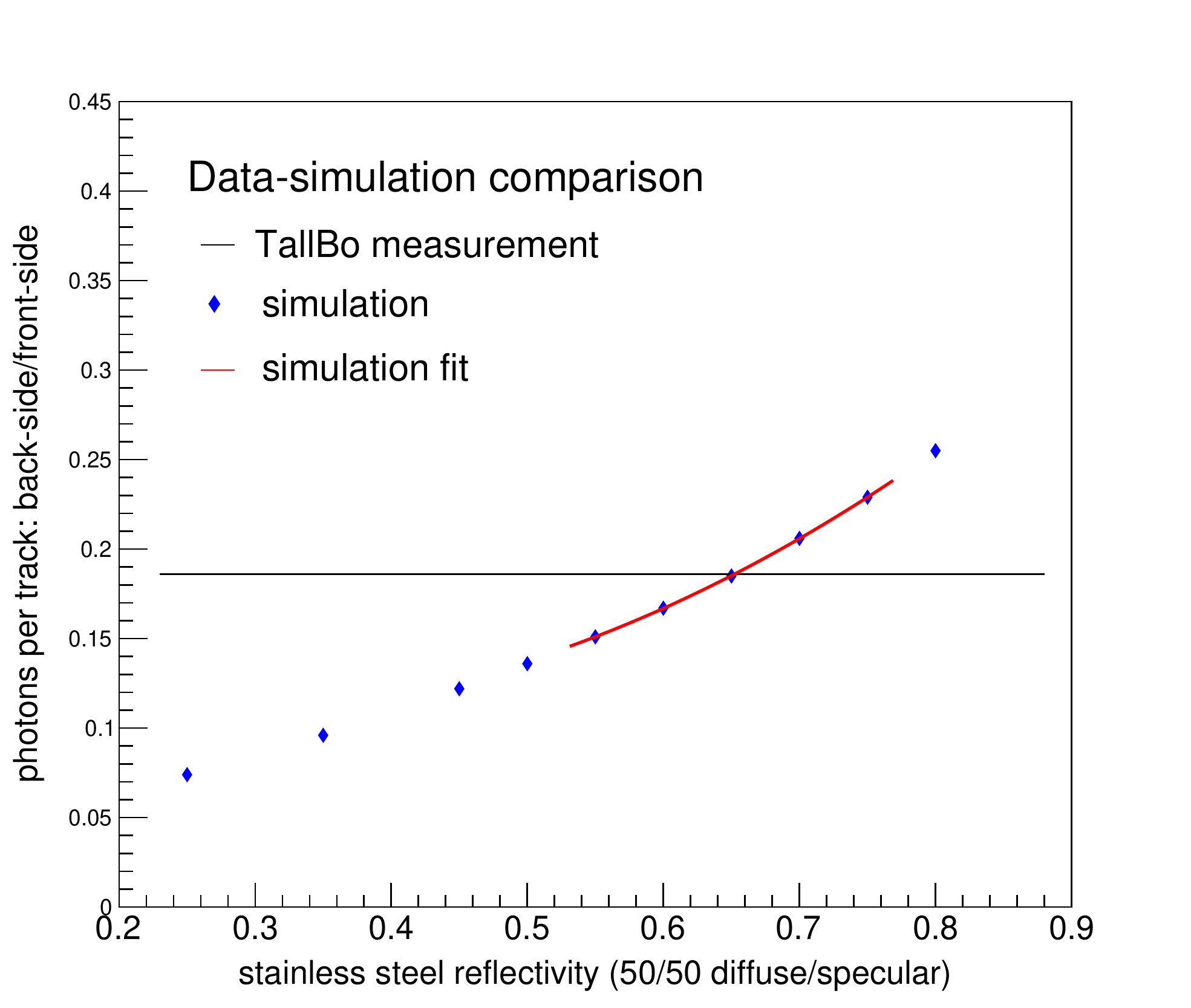}
\caption{The simulations for the stainless steel refectivity between 0.25 and 0.80 superposed on the (run3/run2) ratio from Table~\ref{tab:Results} for MPPC 1 and and Cut (1) = st.dev = Cut (2) = st.dev-long = 3.0.  The ratio of specular to diffuse reflection in these simulations was set to 0.50.  The second order fit to the simulations for the reflectivities 0.55, 0.60, 0.65, 0.70, and 0.75 is also superposed.  The intersection of this fit with the (run3/run2) ratio is the stainless steel reflectivity for MPPC~1 and these analysis cuts.}
\label{money}
\end{figure}

Fig.~\ref{money} shows one set of simulations for MPPC 1 and the stainless steel refectivity between 0.25 and 0.80.  
These simulations are superposed on the (run3/run2) ratio from Table~\ref{tab:Results} for Cut (1) = st.dev = Cut (2) = st.dev-long = 3.0.  In these simulations the ratio of specular to diffuse reflection was  set to 0.50.  Fig.~\ref{money} also shows a second order fit to the simulations for the reflectivities between 0.55 and 0.70.  The intersection of this fit line with the (run3/run2) ratio was taken as the stainless steel reflectivity for MPPC~1 and these analysis cuts.

The procedure described was repeated for all remaining values of the (run3/run2) ratio for MPPC~1.  A similar procedure was employed to find the reflectivities for MPPC~2 and MPPC~3.  The mean of the reflectivities for the four sets of analysis cuts for each MPPC are given in Table~\ref{tab:reflectivity}. 
\begin{table}[ht]
\begin{center}
	\caption{Reflectivities of stainless steel for the ratio of specular to diffuse reflection = 0.50.}
	\vspace{0.2em}
	\label{tab:reflectivity}
	\begin{tabular}{| c |  c |}
		\hline
		\hline
 &  <reflectivity>\\
	\hline
MPPC 1 & ~~0.652$\pm$0.012~~\\  
MPPC 2& ~~0.690$\pm$0.020~~\\  
MPPC3 & ~~0.681$\pm$0.016~~\\  
\hline
\hline
<MPPC> &~~0.674$\pm$0.038~~\\
\hline
\hline
	\end{tabular}
	\end{center}
\end{table}
The error in the mean is the standard deviation of the measurements.  The last row of Table~\ref{tab:reflectivity} gives the mean value of the reflectivity for the 3 MPPCs.  The error in the mean was computed as the quadratic sum of (1) the standard deviation of the three reflectivities used to compute the mean and (2) the standard deviations for the individual means.

\begin{figure}[h]
\centering
\includegraphics[width=3.8in,height=3.2in]{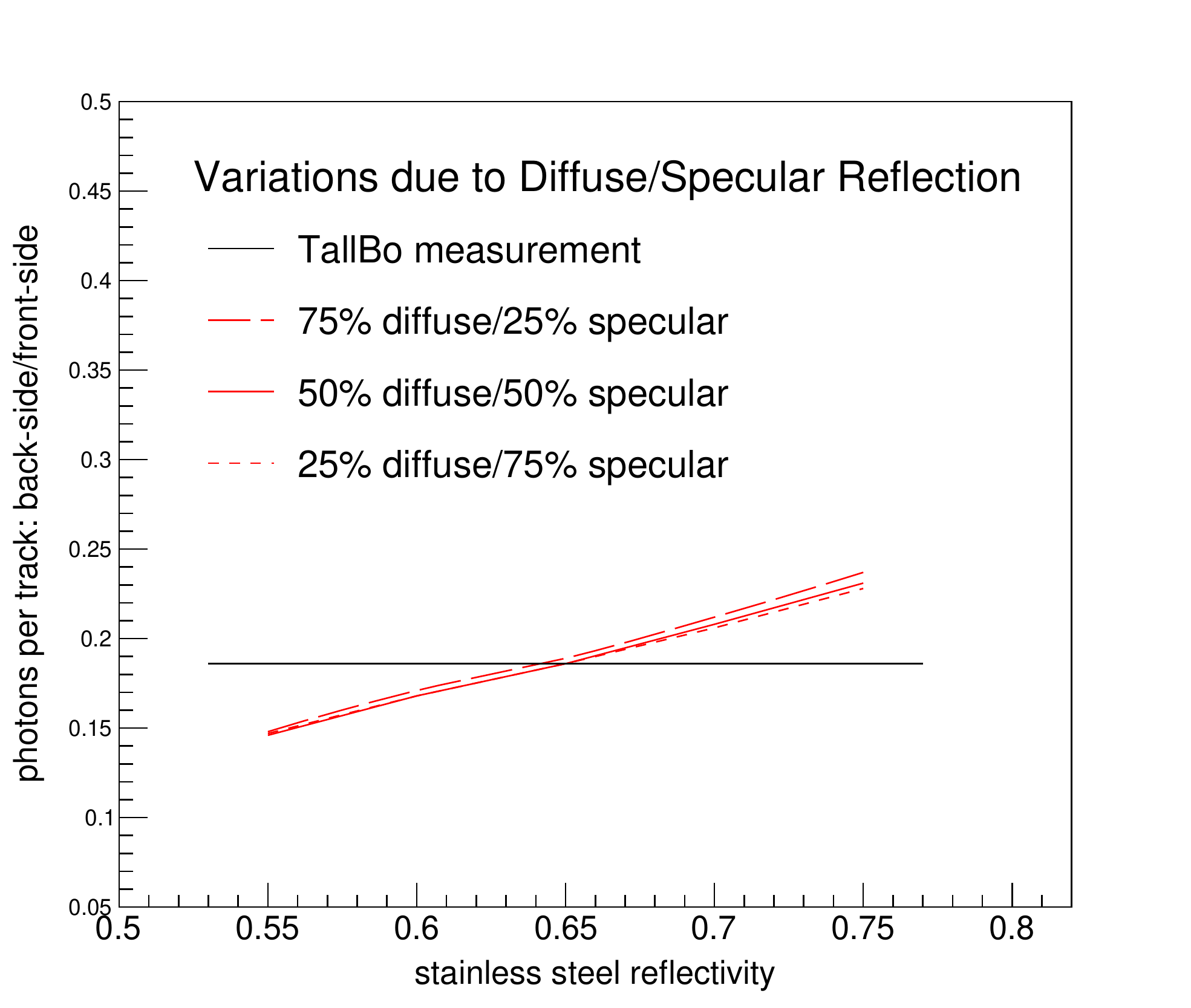}
\caption{Simulations to determine the reflectivity for the same analysis shown in Fig.~\ref{money} for three ratios of the the ratio of specular to diffuse reflection.  The differences in the simulations are small and it is not possible to discriminate their intersections with the (run3/run2) ratio within the systematic errors of the experiment.}
\label{diffuse-specular}
\end{figure}

This experiment is not sensitive to the differences between specular and diffuse reflection as illustrated in Fig.~\ref{diffuse-specular}. 
This figure shows simulations for the same computation for the reflectivity in Fig.~\ref{money} for three ratios of the diffuse-to-specular reflection: 75\%/25\%, 50\%/50\%, and 25\%/75\%.  The reflectivities found for these 3 simulations are 0.639, 0.645, and 0.648, respectively.  The ArCLight detector used in this experiment functioned imperfectly and the small dimensions of the TallBo dewar make this a systematics limited experiment and it is unlikely that longer running time could significantly improve the precision of the (back side/front side) ratios.  Consequently, the differences shown in  Fig.~\ref{diffuse-specular} are undetectable.  

\section{Conclusions}
\label{sec:conclusions}

The value found in this investigation for the reflectivity of the SA-240-304L stainless steel alloy off which the scintillation photons reflect in the TallBo dewar is 
\begin{equation}
R = 0.674 \pm 0.038
\label{finalResult}
\end{equation}
The ratio of diffuse-to-specular reflection cannot be determined by this experiment.

There is an uncertainty in this result, however, that is difficult to quantify.  As discussed above, the ArCLight functioned imperfectly during the experiment. In fact,  a prime purpose of the analysis cuts was to remove anomalous waveforms from the data samples analyzed that can be attributed to these ArCLight performance problems.  There is the possibility, however, that not all anomalous waveforms were identified and removed by these cuts, particularly from the lower signal run~3 data.  Consequently, studies were initiated that looked to eliminate background events in the run~3 waveforms by defining a reflected waveform as one that showed a negative peak after the trigger.  This approach was unsuccessful.  The value of the (run~3/run~2) ratio, and therefore the reflectivity, depended strongly on the choice of the peak strength.  Eliminating tracks in the lowest bins of the integrated waveforms, like those shown in Fig.~\ref{distributions}, proved to be the most objective method for removing anomalous waveforms.  If significantly more background waveforms are present in bins with higher integrated signal strengh in run~3, it might be expected that less restrictive cuts would introduce more of these events in the run~3 data sample and there would be a downward trend in the ratios in Table~\ref{tab:Results}.  This does not seem to be the case.  Therefore, eq.(\ref{finalResult}) appears to be the best choice for the reflectivity of stainless steel in this investigation.


\acknowledgments

\noindent This work was supported in part by the Trustees of Indiana University, the DOE Office of High Energy Physics through grant DE-SC0010120 to Indiana University, and grant \#240296 from Broookhaven National Laboratory to Indiana University.  
The author wishes to thank the many people who helped make this work possible. At Indiana~U.: B.~Adams, B.~Baugh, M.~Gebhard, M.~Lang, C.~Macias, J.~Urheim.  At Fermilab: R.~Davis, A.~Hahn, B.~Howard, B.~Miner, T.~Nichols, E.~Niner, B.~Ramson.  At U.~Wisconsin: B.~Rebel.  At ANL: G.~Drake.  At CERN: I.~Kreslo.

This manuscript has been authored by Fermi Research Alliance, LLC under Contract No. DE-AC02-07CH11359 with the U.S. Department of Energy, Office of Science, Office of High Energy Physics. The United States Government retains and the publisher, by accepting the article for publication, acknowledges that the United States Government retains a non-exclusive, paid-up, irrevocable, world-wide license to publish or reproduce the published form of this manuscript, or allow others to do so, for United States Government purposes.

\hfil

\bibliography{ReflectionsOffStainlessSteel}

\providecommand{\href}[2]{#2}\begingroup\raggedright\begin{thebibliography}{1}

\bibitem{bib:howard}
B.~Howard et~al., {\it A novel use of light guides and wavelength shifting
  plates for the detection of scintillation photons in large liquid argon
  detectors},  {\em NIM} {\bf A907} (2018) 9.

\bibitem{Abi:2020loh}
{\bf DUNE} Collaboration, B.~Abi et~al., {\it {Volume IV. The DUNE far detector
  single-phase technology}},  {\em JINST} {\bf 15} (2020), no.~08 T08010,
  [\href{http://arxiv.org/abs/2002.03010}{{\tt arXiv:2002.03010}}].

\bibitem{bib:IcarusReflection}
M.~Antonello et~al., {\it Analysis of liquid argon scintillation light signals
  with the {ICARUS} {T}600 detector},  {\em Technical Report ICARUS-TM/06-03}
  (2006).

\bibitem{bib:mufson-TallBo}
S.~Mufson et~al., {\it Differences in the response of two light guide
  technologies and two readout technologies after an exchange of liquid argon
  in the dewar},  {\em NIM} {\bf A976} (2020) 164240.

\bibitem{bib:ArCLight}
M.~Auger et~al., {\it {A}r{CL}ight - a compact dielectric large-area photon
  detector},  {\em Instruments} {\bf 2} (2018) 3.

\bibitem{bib:scintYield2}
T.~Doke et~al., {\it Absolute scintillation yields in liquid argon and xenon
  for various particles},  {\em Jpn.J.Appl.Phys.} {\bf 41} (2002) 1538.

\bibitem{bib:TallBo}
D.~Whittington, S.~Mufson, and B.~Howard, {\it Scintillation light from
  cosmic-ray muons in liquid argon},  {\em JINST} {\bf 11} (2016) P05016.

\bibitem{bib:N2Contamination}
R.~Acciarri et~al., {\it Effects of nitrogen contamination in liquid argon},
  {\em JINST} {\bf 5} (2010) P06003.

\end{thebibliography}\endgroup
\bibliographystyle{JHEP}

\end{document}